# Generation and manipulation of chiral terahertz waves emitted from the three-dimensional topological insulator $Bi_2Te_3$

Haihui Zhao,[a,†] Xinhou Chen,[b,†] Chen Ouyang,[c,e,†] Hangtian Wang,[a,d,†] Deyin Kong,[b] Peidi Yang,[b] Baolong Zhang,[c,e] Chun Wang,[c,e] Gaoshuai Wei,[c,e] Tianxiao Nie,[a,d,*] Weisheng Zhao,[a,d,*] Jungang Miao,[b] Yutong Li,[c,e] Li Wang,[c,*] and Xiaojun Wu[b,f,*]

[a]Beihang University, Fert Beijing Institute, BDBC, and School of Microelectronics, Beijing, China, 100191

[b]Beihang University, School of Electronic and Information Engineering, Beijing, China, 100083

[c]Beijing National Laboratory for Condensed Matter Physics, Institute of Physics, Chinese Academy of Sciences, Beijing, China, 100190

[d]Beihang University, Beihang-Goertek Joint Microelectronics Institute, Qingdao Research Institute, Qingdao, China, 266000

[e]University of Chinese Academy of Sciences, School of Physical Sciences, Beijing, China, 100049

[f]Huazhong University of Science and Technology, Wuhan National Laboratory for Optoelectronics, China, 430074

# Abstract

Arbitrary manipulation of broadband terahertz waves with flexible polarization shaping at the source has great potential in expanding real applications such as imaging, information encryption, and all-optically coherent control of terahertz nonlinear phenomena. Topological insulators featuring unique spin-momentum locked surface state have already exhibited very promising prospects in terahertz emission, detection and modulation, which may lay a foundation for future on-chip topological insulator-based terahertz systems. However, polarization shaped terahertz emission with prescribed manners of arbitrarily manipulated temporal evolution of the amplitude and electric-field vector direction based on topological insulators have not yet been explored. Here we systematically investigated the terahertz radiation from topological insulator $Bi_2Te_3$ nanofilms driven by femtosecond laser pulses, and successfully realized the generation of efficient chiral terahertz waves with controllable chirality, ellipticity, and principle axis. The convenient engineering of the chiral terahertz waves was interpreted by photogalvanic effect induced photocurrent, while the linearly polarized terahertz waves originated from linear photogalvanic effect induced shift currents. We believe our works not only help further understanding femtosecond coherent control of ultrafast spin currents in light-matter interaction but also provide an effective way to generate spin-polarized terahertz waves and accelerate the proliferation of twisting the terahertz waves at the source.

*Xiaojun Wu, E-mail: xiaojunwu@buaa.edu.cn
*Tianxiao Nie, E-mail: nietianxiao@buaa.edu.cn
*Weisheng Zhao, E-mail: weisheng.zhao@buaa.edu.cn
*Li Wang, E-mail: wangli@aphy.iphy.ac.cn

# Introduction

Spin-polarized terahertz waves carrying twisted electric and magnetic field directions and temporal evolution of the amplitudes, have been widely utilized for non-thermal and selective excitation of electron spins, resulting in the discovery of spin-galvanic effect[1], the manipulation of terahertz nonlinear phenomena and so on[2-5]. Recently, Kimel's group found that linearly polarized terahertz waves can nonlinearly excite of GHz spin waves in antiferromagnetic $FeBO_3$ via terahertz field induced inverse Cotton-Mouton effect[6]. However, the inverse Faraday effect excited by circularly polarized terahertz waves has not yet been experimentally demonstrated, although many theoretical predictions have been reported[7]. Therefore, spin-polarized terahertz waves are essential for revealing electron spin related linear and nonlinear ultrafast phenomena. Besides, spin-polarized terahertz waves naturally carrying chiral property offers multifaceted spectroscopic capabilities for investigating low-energy macromolecular vibrations in biomaterials, understanding the mesoscale chiral architecture[8], and confidential communication[9,10]. However, the proliferation of terahertz circular dichroism spectroscopy and other applications is impeded by the lack of generation and manipulation of chiral terahertz waves.

Conventional polarization shapers working in terahertz electromagnetic frequency range are mainly segmental waveplate and metasurfaces[11,12], both of which can be well applied for narrow bandwidth applications and low tolerance of flexible tunability. Various scalar manipulation of terahertz waves has been demonstrated in poled nonlinear optical crystals[13,14], or even in water[15] and synthesizing different frequencies generated in spatially distributed sources[16]. However, it would become much more elegant if polarization shaping functionality could be vectorized and integrated into the sources. Up to now, there are already several such kinds of vector manipulated polarized terahertz sources[17-21]. For example, femtosecond laser amplifier pumped air-plasma, when intensively tuning the phase delay as well as the pump pulse polarization between two different pump pulse colors, one can have the opportunity to obtain arbitrarily shaped terahertz polarizations[17]. Furthermore, terahertz polarization pulse shaping with arbitrary field control is also demonstrated by optical rectification of a laser pulse whose instantaneous intensity and polarization state are flexibly controlled by an optical pulse shaper[18]. Recently, elliptically and circularly polarized terahertz waves have been realized in W/CoFeB/Pt heterostructures through deliberately engineering the applied magnetic field distribution and cascade emission method, respectively[19,20]. In NiO single crystal, double pulse excitation has also been used to vectorially control of magnetization by radiating spin-polarized terahertz waves[21], in which electron spin-polarized wave (magnon oscillation) has been purposely vectorially controlled and manipulated. In addition, highly efficient chiral terahertz emission with elliptical polarization states has been obtained from the fascinating topological phase matter, bulk Weyl semimetal TaAs with reflection terahertz emission geometry[22]. Inspired by these pioneering breakthroughs, whether we can utilize the spin freedom of electrons to generate and manipulate spin-polarized terahertz waves has become a very interesting question.

Three-dimensional topological insulator $Bi_2Te_3$ single crystals grown on $Al_2O_3$ substrates, not only feature the spin-momentum locked (topologically protected) surface states but also are adapted for various heterostructures with tunable quantum layers[23]. It is predicted to have the potential for achieving vectorial polarization shaping terahertz sources. The charge current and spin current inside topological insulators are simultaneously excited when being driven by femtosecond laser pulses[24-28]. However, topological insulator-based terahertz emission mechanisms are far more complicated when compared with conventional terahertz emitters. Linear and nonlinear effects have already been, to some extent, clarified that nonlinear effects play a predominant role in the ultrafast emission process[29]. Nonlinear effects include optical rectification, photogalvanic effect and photon drag effect, among which photogalvanic effect has a strong correlation to the electron spin. Thus, there is the feasibility to generate a femtosecond laser helicity dependent current component resulting in synthesizing chiral terahertz light.

In this work, we systematically studied terahertz emission from $Bi_2Te_3$ nanofilms with various thicknesses grown on $Al_2O_3$ substrates excited by femtosecond laser pulses. Through elaborately controlling the pump laser pulse polarization, its incident angle as well as the sample azimuthal angle, we not only successfully realize efficient linear polarization tunable terahertz radiation, but also obtain high quality chiral terahertz waves, as shown in Fig. 1a. Besides, we also demonstrate arbitrarily manipulation of the ellipticity and chirality of the emitted terahertz waves from topological insulators. Based on the radiated terahertz characteristics which originated from the helicity-dependent photocurrent component featuring the spin selectivity in the surface state (Fig. 1b) as well as other component in the bulk state[30] (Fig. 1c), we provided a phenomenological photogalvanic effect (PGE) based interpretation for the polarization tunable terahertz radiation mechanism. Such polarization shaped topological insulator-based terahertz source can be used for terahertz circular dichroism spectroscopy, terahertz secure wireless communication, and electron spin correlated coherent excitation and manipulation investigations.

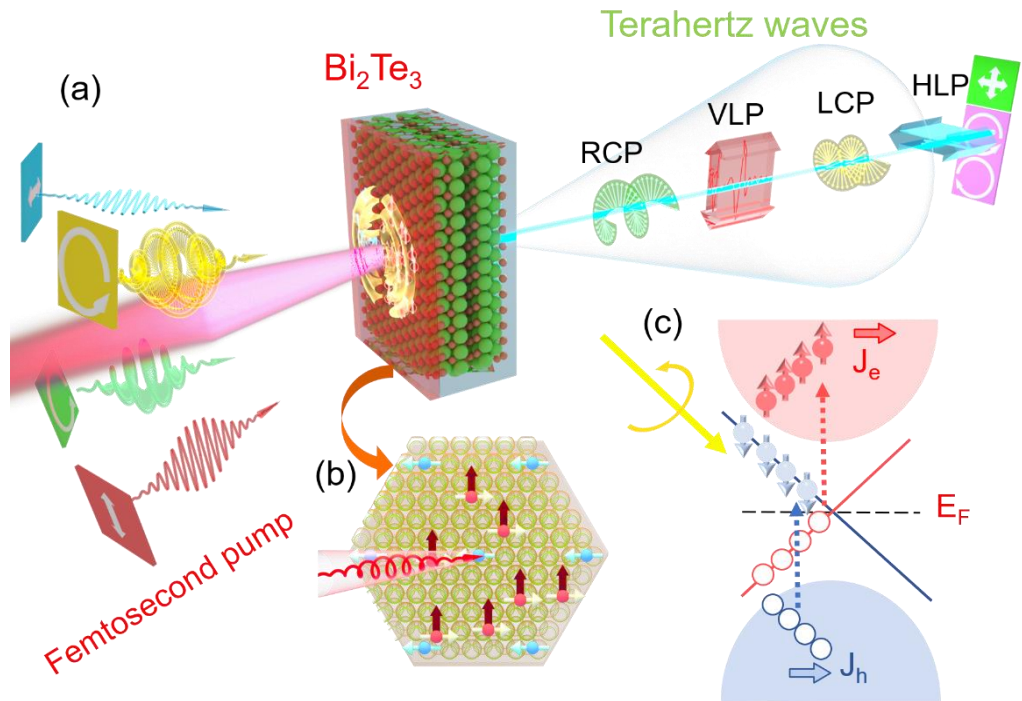

**Fig. 1. Schematic diagram of the concept.** (a), Experimental setup. Horizontal linearly polarized (HLP), vertical linearly polarized (VLP), left-handed circularly polarized (LCP) and right-handed circularly polarized (RCP) femtosecond laser pulses are incident onto the topological insulator and produce corresponding polarized shaped terahertz waves. (b), Macroscopic helicity-dependent photocurrent and only unidirectional spin current can be generated. (c), Microscopic electronic transition under circularly polarized laser pulse illumination.

# Linearly polarized terahertz emission

In this experiment, we employed a commercial Ti:sapphire laser oscillator delivering nJ magnitude pulse energy with a central wavelength of 800 nm, a pulse duration of 100 fs, and a repetition rate of 80 MHz. The pump laser polarization was varied by inserting either a half wave plate or a quarter wave plate. The radiated terahertz polarization was probed by a polarization-resolved electro-optic sampling detection method. The topological insulator samples were grown by molecular beam epitaxy (MBE) and their lattice structures as well as the characterizations can be found in Fig. S1.

At first, we examined the sample azimuthal angle dependent terahertz radiation polarization properties to unveil radiation mechanism in topological insulators driven by femtosecond laser pulses. For p-polarized pump laser at nearly normal incidence, as shown in Fig. 2a, the radiated terahertz signal from 10 nm thick $Bi_2Te_3$ was ~1/20 compared with that from 1 mm thick ZnTe (Fig. S2), and with ~100 dynamic range, which manifests its feasibility for terahertz spectroscopy. The radiated terahertz waves were always linearly polarized with a three-fold rotation angle depending on the sample azimuthal angle, as shown in Fig. 2b and c. For the pump-fluence-dependence results (Fig. 2d), the terahertz signals of different thickness all exhibited a linear increasing tendency, indicating the terahertz radiation mechanism under the linear-polarization laser pump was predominant by a second-order nonlinear effect.

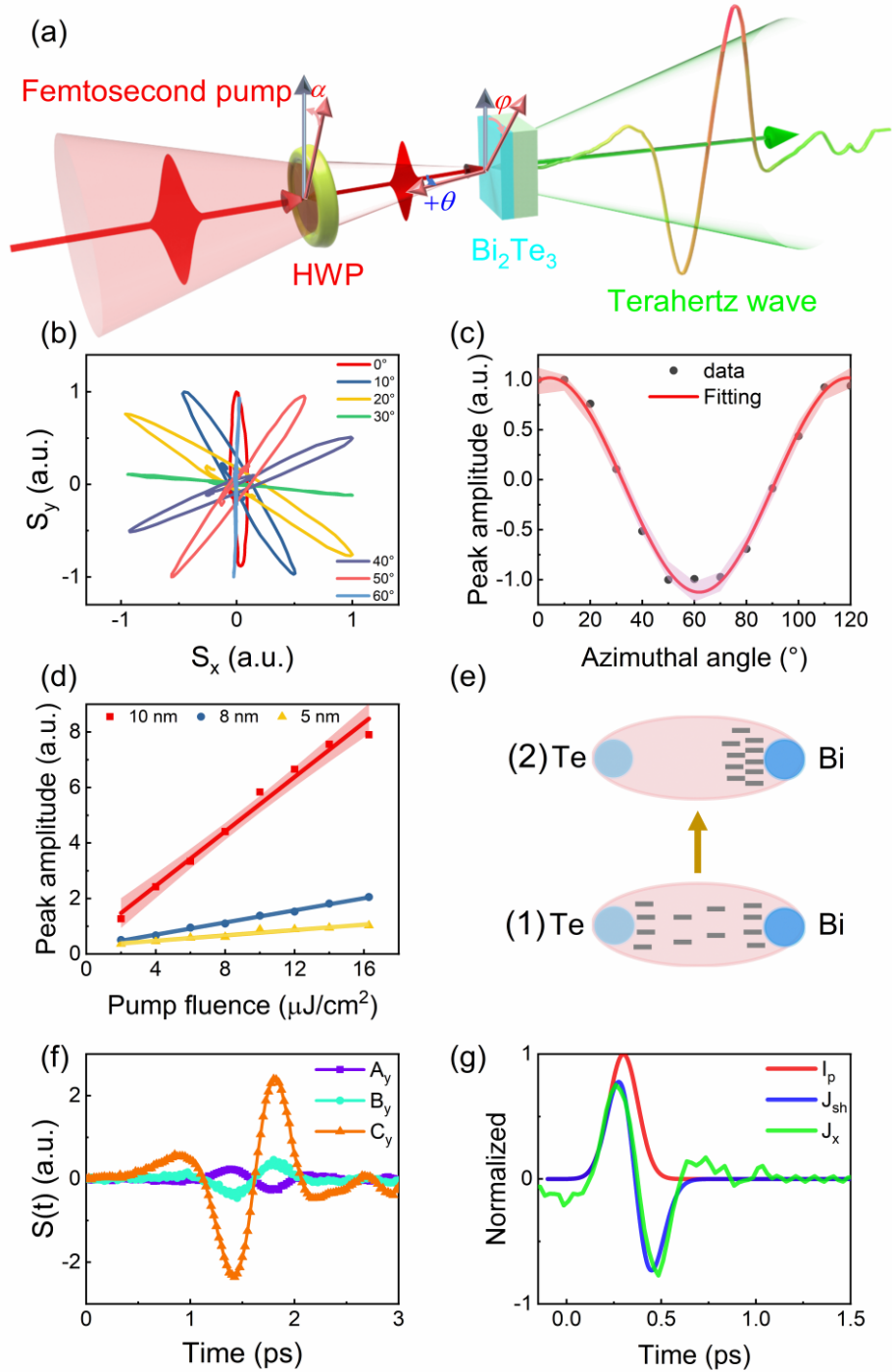

**Fig. 2. Linearly polarized terahertz emission and its shift current mechanism.** (a), Schematic diagram of the experimental setup for linearly polarized terahertz emission. Linearly polarized laser pump passes through a half wave plate of angle $\alpha$, and incident onto the sample emitting linearly polarized terahertz wave. The incident angle of the pump laser: $\theta$; the azimuthal angle of the topological insulator: $\varphi$. (b), Far-field polarization trajectories ($S_x(t)$,

$S_y(t)$) of the radiated terahertz waves obtained experimentally from 10 nm thick $Bi_2Te_3$ when the pump laser polarization was nearly normal incidence while the sample azimuthal angle was varied. (c), The terahertz peak amplitude of $S_y(t)$ as a function of the azimuthal angle exhibits a 120° period. (d), Pump fluence dependence of terahertz peak amplitude from $Bi_2Te_3$ with different thickness of 5 nm, 8 nm, and 10 nm. (e), Schematic of the shift current generating from the electron transfer along the Bi-Te bonds. (f), Parameters in y-axis extracted through the symmetry analysis of surface state using Eq. (1). (g), Fitting the shape of shift current (blue line) to the photocurrent in the material (green line).

Nonlinear effect radiation mechanisms include optical rectification (OR), photon-drag effect (PDE) and photogalvanic effect (PGE). To evaluate the relative importance of optical rectification in our experiment, we compared the emitted terahertz amplitude from the topological insulator sample with that from a ZnTe crystal, which is known for relatively strong optical rectification at the pump wavelength used here. We found that the terahertz signal from topological insulator was almost two orders of magnitude larger than that from ZnTe when we normalized the amplitudes by the emitter thicknesses. Therefore, we can safely rule out optical rectification as the predominant radiation mechanism[31]. Linear photon-drag effect (LPDE) can be microscopically understood as the linear transfer from photon momentum to electron momentum, which means when the opposite incident angle was employed, we would obtain opposite polarity of terahertz signals. However, when we varied the incident angles with opposite signs under the p-polarized laser pumping, which was a pure LPDE case, such behaviors neither for $S_x(t)$ nor for $S_y(t)$ were not observed, as depicted in Fig. S4. Hence, we can further safely remove the LPDE as the primary radiation origin.

Macroscopically, PGE is a second-order nonlinear process occurring in the noncentral symmetric structure. From microscopic physical picture, PGE can be understood as optical transition on the surface states[31,32], which can be divided into linear PGE (LPGE) and circular PGE (CPGE). The linear photogalvanic current, also named as a shift current, can be explained that the electron density distribution is spatially shifted from the state (1) to state (2) along the Bi-Te bonds (see Fig. 2e) when the pump laser pulses illuminate onto the topological insulator, following with relaxation process such as asymmetric scattering due to the optical electric field[31]. For this mechanism underlying the terahertz emission scenario, we can quantify the behavior on both x- and y-direction by analyzing the symmetry of the terahertz signals. The three-fold rotational symmetry of signals is in coordinate with the space group of the surface and bulk such that we can write the two-dimensional waveform set $S(t,\varphi)$ as a linear combination of three basis[31],

$$S(t,\varphi) = A(t) + B(t)\sin(3\varphi) + C(t)\cos(3\varphi) \tag{1}$$

Thus, these three basis can completely characterize the $S(t,\varphi)$. Among the three coefficients, $B(t)$ and $C(t)$ represent the pure LPGE-dependent components, while $A(t)$ represents a direct current component which might be ascribed to a thermal current caused by the thermal potential gradient when the sample was under pumping. From the extracted coefficients in y-axis as shown in Fig. 2f (coefficients in x-axis can be found in Fig. S5), it is obvious that the thermal current contributes much less than the shift current. Therefore, the thermal effect in our Bi$_2$Te$_3$ material under excitation can be safely ignored. We only need to consider the shift current in the generation process of linearly polarized terahertz wave scenario.

As for a more profound insight about the shift current generated under the short pulse excitation, this process results in a step-like charge displacement $\Delta x_{sh} \Theta(t)$, where $\Delta x_{sh}$ is the spatial displacement of the electron density, $\Theta(t)$ denotes the unit step function, whose temporal derivative is proportional to the shift current $J_{sh}$. A phenomenological relaxation process such as scattering also should be included so that an exponential decay term with time constant $\tau_{sh}$ has to be introduced. Therefore, we obtain the shape of the shift current as follows,

$$J_{sh} \propto \Delta x_{sh} \frac{\partial}{\partial t}\left[\Theta(t)\exp\left(-\frac{t}{\tau_{sh}}\right)\right] * I_p \qquad (2)$$

The convolution (marked by $*$) with the laser intensity profile $I_p(t)$ (normalized to unit) accounts for the influence under the process of excitation. This model illustrates that the profile of $J_{sh}$ initially follows the envelope of $I_p(t)$ and subsequently becomes bipolar as shown in Fig. 2g (blue line). Due to the relatively flat response function of ZnTe[33] within the low frequency range (<3 THz), we can semi-quantitatively retrieve the photocurrent (green line) inside the material (see Methods and Fig. S6). Moreover, in order to fit the photocurrent through the theoretical model, the $I_p(t)$ unambiguously exhibits an extension of its pulse duration to 140-170 fs (red line), and the retrieved photocurrent agrees well with the calculated results among the whole profile. According to the fitting process, we obtain the relaxation time $\tau_{sh} = 22$ fs, which is consistent with that in the previous literature[31], further proving the validity of LPGE mechanism under this circumstance. Note that, it has been reported the shift current in Bi$_2$Te$_3$ originated from the transient electron transfer along the Bi-Te bonds involving the surface states related optical transitions. However, for the ultrathin $Bi_2Te_3$ nanofilms, the coupling between the top and bottom surface opens a gap in the surface state dispersion, and hence suppresses the surface states related optical transitions. Therefore, as shown in Fig. 2d, the shift current as well as its resulting terahertz emission decreases considerably.

## Chiral terahertz wave emission and manipulation

As shown in Fig. 3a, when the left-handed circularly polarized pump light was incident onto the sample at an incident angle around +20°, and we rotated the sample azimuthal angle, elliptically polarized terahertz waves were obtained and showed no significant change when scanning the azimuthal angle $\varphi$ in Fig. 3b. To produce perfect circularly polarized terahertz pulses, simultaneously tuning the pump laser polarization as well as the sample azimuthal angle was necessary. The experimental results are exhibited in Fig. 3c. When the sample azimuthal angle was fixed, we can also obtain elliptical terahertz beams with various ellipticities and azimuthal angles. Within expectation, we can manipulate the chirality of the emitted terahertz waves via varying the incident laser helicity. It's obvious that the polarity of emitted terahertz signal along x-direction reversed under illumination of laser with different chirality (Fig. 3d), and sustain the sign along the y-z plane (Fig. 3e). In short, this is the critical reason why we can obtain spin-polarized terahertz pulse with high quality.

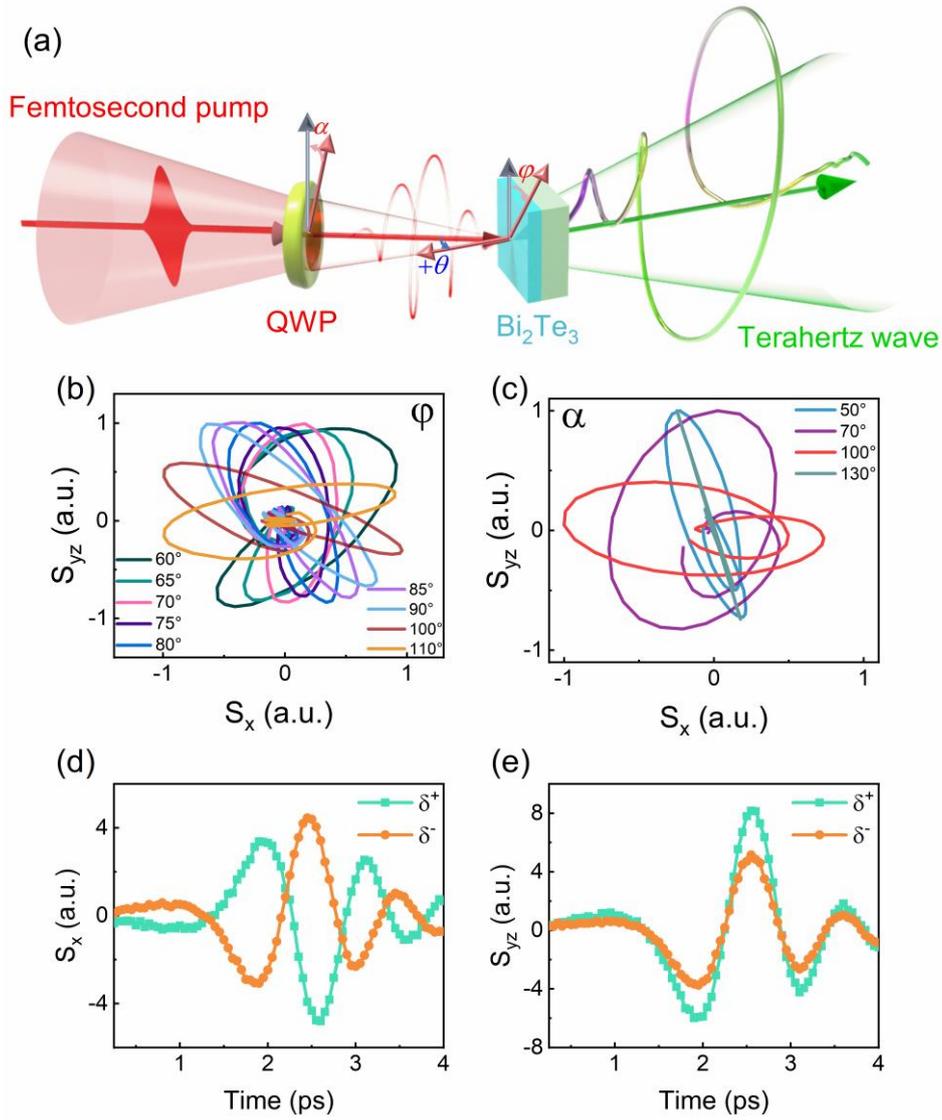

**Fig. 3. Generation of elliptically and circularly polarized terahertz beams.** (a), Experimental layout for circular terahertz wave generation. Linearly polarized laser pump passes through a quarter wave plate of angle $\alpha$, producing elliptically or circularly polarized pump laser beams. The emitted terahertz wave polarization can be elliptical and circular. The incident angle of the pump laser: $\theta$; the azimuthal angle of the topological insulator: $\varphi$. (b), Far-field detected elliptically polarized terahertz polarization trajectories when the pump laser polarization was fixed while rotating the sample azimuthal angles. (c), Production of circularly polarized terahertz waves when fixing the azimuthal angle while rotating the quarter wave plate

for the pump laser pulses. (d), Experimentally observed terahertz component $S_x(t)$ polarity reversal depends on the pump laser helicity. (e), Helicity-independent component $S_{yz}(t)$ from Bi$_2$Te$_3$. $\delta^+$ and $\delta^-$ mean left-handed and right-handed elliptically polarized terahertz, respectively.

In order to further investigate the helicity dependent terahertz signal, a terahertz pulse with chirality was manipulated by rotating the quarter wave plate with an angle $\alpha$ at a fixed incident angle around +20°. When proceeding the $\alpha$ scanning, the terahertz peak values clearly exhibited a 180° period (Fig. 4a), and the waveform shape along x-axis was very similar to the result in Ref. 24 in Sb$_2$Te$_3$ thin film system. It illustrates the analogy among the family of topological insulators.

Therefore, based on the geometric structure of the incident laser which was modulated via the quarter wave plate (Fig. S8), we can write the signal in the following form[25,26,34],

$$S(t,\alpha) = C(t)\sin(2\alpha) + L_1(t)\sin(4\alpha) + L_2(t)\cos(4\alpha) + D(t) \qquad (3)$$

The coefficient $C(t)$ describes the helicity dependent terahertz radiation which originates from CPGE. $L_1(t)$ denotes the coefficient induced by linearly polarized light, and LPGE may be responsible for it. For terahertz emission in 40 nm Sb$_2$Te$_3$, Yu et, al[34] attributed the origin of $L_2(t)$ to LPDE, and experimentally verified the trend of $L_2(t)+D(t)$ under the variation of the incidence angle $\theta$ was in accordance with the trend of pure LPDE for p-polarized laser excitation. However, as aforementioned, LPDE has already been eliminated so it isn't suitable for elucidating this phenomenon. In terms of $D(t)$, there isn't a uniform statement, and is still under controversial debates. LPGE, drift current out of plane or Photo-Dember effect may be the candidates for this coefficient.

Accordingly, we extracted all coefficients along x-axis and y-z plane (see Methods). All data were offset to ensure great clarity as shown in Fig. 4b and c, which show the time-domain trace for every component. The main characteristics are summarized as follow: (*i*) In magnitude, along the x-axis, $C_x$, $L_{1x}$ and $D_x$ dominate the main components while $L_{2x}$ is negligible. They are sorted from the largest to smallest $L_{1x} > D_x > C_x$. Along y-z plane, $L_{2yz}$ and $D_{yz}$ dominate the main components, while $L_{1yz}$ and $C_{yz}$ are negligible. In addition, $L_{2yz}$ and $D_{yz}$ have almost the same magnitude. (*ii*) $C_x$ shows similar characteristics to those for $L_{1x}$. In fast Fourier transformed spectra of $C_x$ and $L_{1x}$ (Fig. 4e and f), they both presented the same spectral shapes strongly indicating that both $C_x$ and $L_{1x}$ shared the same physical origin and are consistent with previous observation[25,26,35]. However, the $D_x$, $L_{2yz}$ and $D_{yz}$ terms also presented similar shapes in the frequency domain, and very few work payed attention to the underlying mechanism about these three terms. It is the first time to observe that all five terms resemble each other in nature, and we will elucidate this phenomenon in a macroscopic phenomenonlogical PGE induced photocurrent framwork.

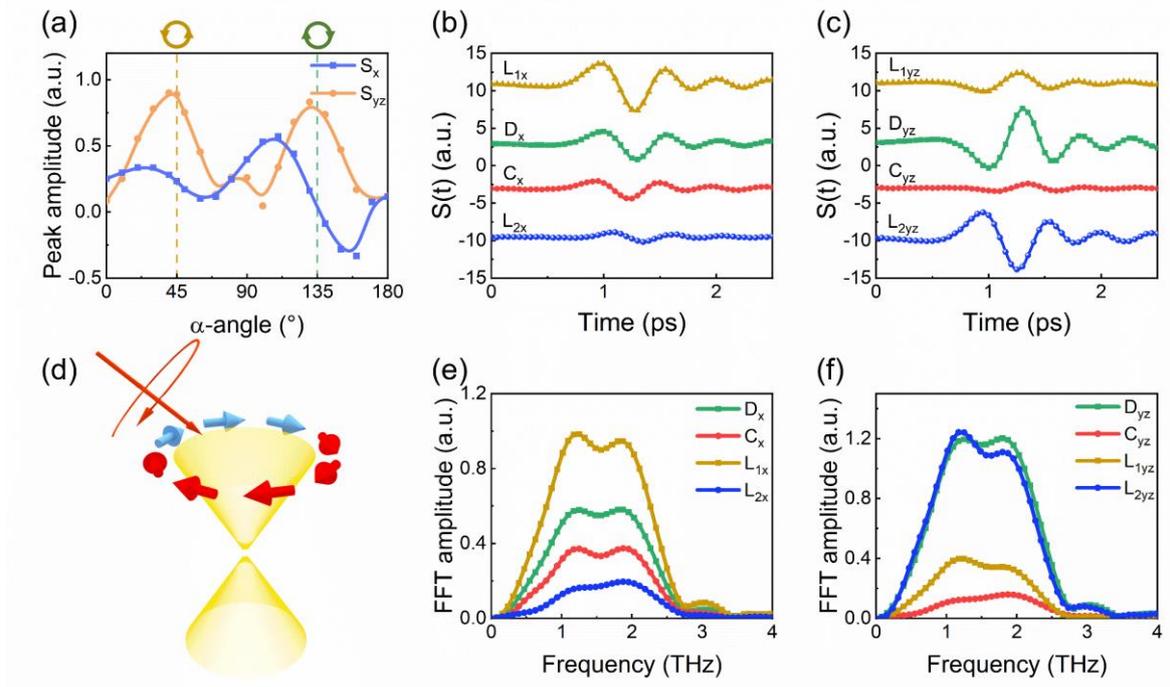

**Fig. 4. Macroscopic analysis of photogalvanic effect.** (a), The terahertz peak amplitude of $S_x(t)$ and $S_{yz}(t)$, as a function of the quarter wave plate angle, respectively. (b) and (c), The time-domain signals for the parameters $C(t)$, $L_1(t)$, $L_2(t)$, and $D(t)$ in x- and y-z plane extracted using Eq. (3). (d), Spin-momentum locked states selectively excited by spin-polarized pump laser form unidirectional spin currents. (e) and (f), The corresponding Fourier transformed spectra of the time-domain signals in (b) and (c).

Macroscopically, as a second-order nonlinear optical process, PGE induced photocurrent vanishes in the bulk region where inversion symmetry exists. Only on the surface where inversion symmetry is broken, the photocurrent induced by PGE can occurr. Thus, taking the $C_{3v}$ symmetry into account, in our specific case in which the incident pulse shines on y-z plane, the arbitary polarized laser pulse induced photocurrent can be seperated to CPGE and LPGE induced components (See the derivation in Supplememtary Note 6):

$$j_{CPGE} = -2\gamma C^2 \begin{pmatrix} \sin 2\alpha \sin\theta \\ 0 \\ 0 \end{pmatrix}, C = \frac{\omega A_0}{2} \tag{4}$$

And the LPGE induced photocurrent can be written in a compact way, $j_{LPGE} = \begin{pmatrix} j_x \\ j_y \\ j_z \end{pmatrix}$, all components are written as

$$j_x = -2C^2 (\eta_1 \cos\theta + \eta_2 \sin\theta) \sin 4\alpha \tag{5}$$

$$j_y = 2C^2 \left[\eta_1\left(\frac{1}{2}\cos^2\theta + \frac{1}{2}\right) - \eta_2 \sin\theta\cos\theta\right] \cos 4\alpha + 2C^2 \left[\eta_1\left(\frac{3}{2}\cos^2\theta - \frac{1}{2}\right) - 3\eta_2 \sin\theta\cos\theta\right] \tag{6}$$

$$j_z = 2C^2 \left[\eta_3 \frac{1}{2}\sin^2\theta + \eta_4\left(\frac{1}{2}\cos^2\theta - \frac{1}{2}\right)\right] \cos 4\alpha + 2C^2 \left[\eta_3 \frac{3}{2}\sin^2\theta + \eta_4\left(\frac{1}{2} + \frac{3}{2}\cos^2\theta\right)\right] \tag{7}$$

where the incident angle is $\theta$, the rotated angle of quarter wave plate is $\alpha$, $\gamma$ and $\eta$ denote the independent components for CPGE and LPGE tensor elements, respectively.

From the above four PGE induced photocurrent equations, we can obtain four conclusions. (*i*) For incident laser with arbitrary polarization and directions, besides the CPGE induced term $j_{CPGE} \propto \sin 2\alpha$, the LPGE induced photocurrent can be written in a form of $j_{LPGE} = j_{LPGE,1} \sin 4\alpha + j_{LPGE,2} \cos 4\alpha + j_{LPGE,3}$. The combination of these two photocurrents explains why we can write the emitted terahertz signals in a form of Eq. (3). (*ii*) We can readily find that when the incident laser pulse changes chirality from left-handed ($\alpha = \pi/4$) to right-handed ($\alpha = 3\pi/4$), the CPGE induced photocurrent will change sign. Furthermore, the CPGE induced photocurrent only flows along the direction perpendicular to the incident plane, which is consistent with the results in Fig. 3d. (*iii*) From these four equations, we can qualitatively analyze the magnitude relationship of the photocurrent components. As in the x direction, from Eq. (4) and Eq. (5), only CPGE induced $C_x$ and LPGE induced $L_{1x}$ components contribute to the photocurrent. Because the incident angle is relatively small and $L_{1x}$ contains $\cos\theta$ term, $L_{1x}$ should be significantly larger than $C_x$. It is obvious that there is no $L_{2x}$ contribution in the x direction. In the y-z plane, $C_{yz}$ and $L_{1yz}$ are both 0, and the magnitude of $L_{2yz}$ and $D_{yz}$ are approximately identical. The magnitudes of these current components determined by the crystal symmetry in the x-axis and y-z plane are in good agreement with the relative magnitudes of the various components extracted in our experiments except for $D_x$. There is a relatively large $D_x$, observed in experiment, and the physical origin of this component may come from dark current[30] which needs further investigation. (*iv*) It is worth noting that the PGE theoretical framework illustrate that photocurrent components all originate from the same PGE mechanism which is exactly the evidence reflected by their frequency spectral shapes. Therefore, we can conclude the phenomenological PGE mechanism framework determined by the symmetry of the crystal surface is adequate to explain most of the salient features of the resulting photocurrents.

In order to acquire spin-polarized terahertz pulse with higher quality, we can also readily gain inspiration from the above framework. When the incident laser pulse is left-handed ($\alpha = \pi/4$), $j_{CPGE,x} = -2\gamma C^2 \sin\theta$, $j_{LPGE,x} = 0$, $j_{LPGE,yz} = f(\theta)$. When the incident laser pulse is right-handed ($\alpha = 3\pi/4$), $j_{CPGE,x} = 2\gamma C^2 \sin\theta$, $j_{LPGE,x} = 0$, $j_{LPGE,yz} = f(\theta)$. It's certainly clear that when we change the chirality of laser pulse, we can manipulate the polarity of CPGE induced photocurrent, remaining the orthogonal component invariant. Based on this statement, we can further analyze the manipulation of phase and amplitude for terahertz radiation. As for the phase between x-axis and y-z plane, there actually exist two different physical processes. The physical process in LPGE represents a shift current along Bi-Te bond. As for the process in

CPGE, the laser with certain chirality will selectively excite a certain spin alongside the direction of incidence due to the spin texture of the surface states (Fig. 4d). Subsequently, the asymmetry distribution of spin would result in a photocurrent which is perpendicular to the direction of spin because of the spin-momentum locking[26]. As long as there exists a phase difference between CPGE and LPGE induced photocurrents, which has already been assured by the above two different physical processes, the spin-polarized terahertz radiation is guaranteed. Regarding the amplitude, it's obvious that random choosing incident angle $\theta$ cannot warrant indentical amplitude along x-axis and y-z plane. This explains why we cannot achieve optimal ellipticity when employing circular polarized pulse. Therefore, to optimize the ellipticity, we can increase the incident angle as well and look forward to achieving higher ellipticity in a wider spectrum.

In short, we systematically analyzed the features of $L_1$, $L_2$, $D$ via the symmetry on the surface of $Bi_2Te_3$ under a phenomenological PGE framework for the first time. We realized that the invariant $L_2$, $D$ components and polarity-controllable $C$ component were all prerequisites for the spin-polarized terahertz emission. Furthermore, this framework can summarize most features of extracted components except for $D_x$ and gives hints for achieving optimal spin-polarized terahertz emission, which owns significance of guidance.

Based on the above theoretical analysis, we summarize the terahertz emission experimental results in Fig. 5. From these results, we can see that, for a fixed laser incidence angle, linearly polarized pump light can only produce linearly polarized terahertz waves (Fig. 2 and Fig. 5c), circularly polarized light can produce chiral terahertz waves, and its chirality is consistence with that of the pump laser (Fig. 5a and b). To make the terahertz generation and manipulation more advanced, we can successfully realize various complex terahertz wave polarization shaping in topological insulators.

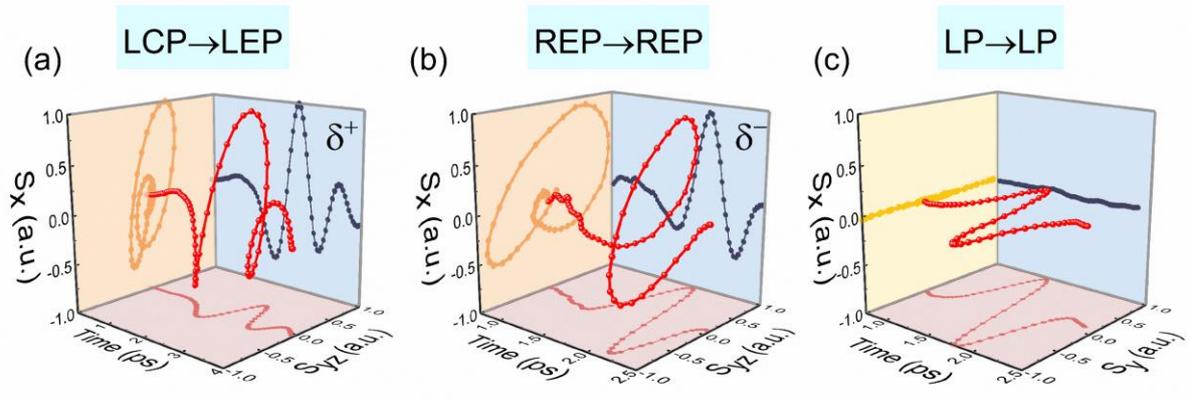

Fig. 5. **Arbitrary manipulation of various terahertz polarization states.** Radiated terahertz chirality was consistence with that of the pump laser chirality. (a), Left-handed circularly polarized pump laser pulses can produce left-handed elliptically polarized terahertz waves, and (b), right-handed elliptically polarized pump light can generate right-handed elliptically polarized terahertz beams. (c), Linearly polarized terahertz waves can be generated in $Bi_2Te_3$ pumped by linearly polarized pump laser beams.

## Conclusion

In summary, we systematically studied terahertz emission from topological insulator $Bi_2Te_3$ nanofilms driven by femtosecond laser oscillator pulses. Through experimentally investigating the sample thickness, azimuthal angle and pump laser polarization dependent terahertz radiation properties, we clarified that photogalvanic effect played a predominant radiation mechanism in this light-matter ultrafast interaction process and producing efficient terahertz waves. Borrowing the spin-momentum locked state induced helicity dependent spin-polarized current generation and combing it with the helicity independent photocurrent, we not only realized circularly polarized terahertz beam generation but also demonstrated arbitrary manipulation of their polarization shaping in topological insulators via simultaneously adjusting the pump laser polarization states and sample azimuthal angles. Our method may have the feasibility to be extended to a variety of topological insulator materials and other related Dirac quantum materials. The topological insulator nanofilms can be grown with high quality and large size possible for high field terahertz sources at transmission emission geometry, which may also provide very promising applications in fundamental nonlinear terahertz investigations and other real applications such as terahertz circular dichroism spectroscopy, polarization-based imaging, and terahertz secure communication.

## Competing financial interests

The authors declare no competing financial interests.

## Acknowledgments

**Funding:** This work is supported by Beijing Natural Science Foundation (4194083), the National Natural Science Foundation of China (61905007, 61774013, 11827807), the National Key R&D Program of China (2019YFB2203102, 2018YFB0407602), the International Collaboration Project (B16001), and the National Key Technology Program of China (2017ZX01032101).

**Author contributions:** X.J.W. and T.X.N. conceived and coordinate the spintronic terahertz emission and ultrafast spin dynamics project. X.J.W. conceived the chiral terahertz wave generation and manipulation idea. H.H.Z., X.H.C. and C.W. carried out the experiments, collected and analyzed the data with help from X.J.W. H.T.W. and T.X.N. fabricated the samples. The theoretical formalisms were derived by C.O. With contributions from M.Z., G.S.W., W.S.Z., J.G.M., Y.T.L., and L.W. contributed with helpful discussions on the experimental and theoretical results. H.H.Z., C.O., X.H.C., M.Z., and X.J.W. wrote the manuscript with revisions by all.

## Methods

**Sample preparation and characterization**

Normal three-dimensional topological insulators are V-VI component semiconductors with hexagonal crystal structure, in which the building block is a quintuple layer (QL). Our topological insulator nanofilms with 10 nm, 8 nm and 5 nm thicknesses grown on 0.5 mm thick sapphire are $Bi_2Te_3$ prepared by molecular beam epitaxy. As shown in Supplementary Fig. 1a, each QL of $Bi_2Te_3$ consists of five atomic layers, terminating by Te atoms without dangling bonds, and a weak interaction called Van der Waals force binds two neighboring QLs. This peculiar gap makes $Bi_2Te_3$ to overcome the large lattice mismatch with sapphire substrates (~8%). Due to the weak adhesion between $Bi_2Te_3$ and sapphire, we employed a two-step deposition procedure. The initial 1-2 QLs of $Bi_2Te_3$ were deposited at low temperature of 160 °C. The lower temperature diminished the material quality which promoted the stacking of atoms to the substrates. After finishing this first deposition, the substrate temperature slowly increased to 230 °C. Based these seed layers, the rest of the film could be deposited with a good quality.

**Terahertz emission spectroscopy**

In our experiment, we used a home-made terahertz time-domain spectrometer to carry out all the measurements. The femtosecond laser pulses were from a commercial Ti:sapphire femtosecond laser oscillator with a central wavelength of 800 nm, a pulse duration of 100 fs, and a repetition rate of 80 MHz. 90% of the laser energy was used for terahertz generation in topological insulators and the residual was employed for electro-optic sampling in 1 mm thick ZnTe detection crystal. Either a half-wave plate or a quarter-wave plate was inserted into the pumping beam before it illuminated onto the sample to vary the pump laser polarization states. The radiated terahertz pulses went through four parabolic mirrors and finally were focused together with the probing beam into the ZnTe detector. Between the second and the third parabolic mirrors, there were two terahertz polarizers inserted into the optical path to resolve the terahertz polarization[18]. The optical path from the terahertz emitter to the detector was sealed and pumped to eliminate the influence of water vapor. All the experiments were carried out at room temperature.

**Extracting coefficients in linear and circular polarized terahertz signals**

According to equation (1), the basis functions $A(t)$, $B(t)$ and $C(t)$ are obtained by multiplying $S(t,\varphi)$ with $3/2\pi$, $3\sin(3\varphi)/\pi$, $3\cos(3\varphi)/\pi$, respectively, and then subsequent integration from $\varphi=0$ to $2\pi/3$. According to equation (3), the basis functions $C(t)$, $L_1(t)$, $L_2(t)$ and $D(t)$ are gained by multiplying $S(t,\alpha)$ with $2\sin(2\alpha)/\pi$, $2\sin(4\alpha)/\pi$, $2\cos(4\alpha)/\pi$ and $1/\pi$ respectively, and then subsequent integration from $\alpha=0$ to $\pi$.

**Retrieving femtosecond photocurrents**

The measured electro-optic signal $S(t)$ and the terahertz electric field $E(t)$ near the sample surface have a linear relationship in the frequency domain as the following equation

$$S(\omega) = H(\omega) E(\omega) \tag{8}$$

where, $H(\omega)$ is the transfer function. After getting the frequency domain $E(\omega)$, the Ohm's law can be used to obtain the photocurrent $J(t)$ on the surface of $Bi_2Te_3$,

$$J_x(\omega) = -\frac{\cos\theta + \sqrt{n^2 - \sin^2\theta}}{Z_0} E_x(\omega) \tag{9}$$

$$J_{yz}(\omega) = -\frac{n^2 \cos\theta + \sqrt{n^2 - \sin^2\theta}}{Z_0 \sqrt{n^2 - \sin^2\theta}} E_{yz}(\omega) \tag{10}$$

Here, $\theta$ is the incident angle, $n$ is the index refractive index of $Bi_2Te_3$ (shown in Supplementary Fig. 6), $Z_0 \approx 377\Omega$ is the vacuum impedance. The photocurrent $J(t)$ can be gained from the inverse Fourier transformation of the current spectra $J(\omega)$. More shapes of the physical parameters are shown in Supplementary Fig. 6.